\documentclass{JAC2000}
\usepackage{graphicx}
\newcommand{\ud}{\mathrm{d}}
\begin{document}
\title{MULTIRESOLUTION REPRESENTATION FOR ORBITAL DYNAMICS IN
MULTIPOLAR FIELDS}
\author{A. Fedorova,  M. Zeitlin, 
IPME, RAS, V.O. Bolshoj pr., 61, 199178, St.~Petersburg, Russia 
\thanks{e-mail: zeitlin@math.ipme.ru}\thanks{ http://www.ipme.ru/zeitlin.html;
http://www.ipme.nw.ru/zeitlin.html}
}

\maketitle

\begin{abstract}
We present the applications of variation --
wavelet analysis to polynomial/rational approximations for
orbital motion in transverse plane for a single
particle in a circular magnetic lattice in case when we take into account
multipolar expansion up to an arbitrary finite number and additional
kick terms. We reduce initial
dynamical problem to the finite number (equal to the number of n-poles) 
of standard
algebraical problems.
We have the solution as
a multiresolution (multiscales) expansion in the base of compactly
supported wavelet basis.
\end{abstract}

\section{INTRODUCTION}

In this paper we consider the applications of a new nu\-me\-ri\-cal-analytical 
technique which is based on the methods of local nonlinear harmonic
analysis or wavelet analysis to the 
orbital motion in transverse plane for a single
particle in a circular magnetic lattice in case when we take into account
multipolar expansion up to an arbitrary finite number and additional
kick terms. We reduce initial
dynamical problem to the finite number (equal to the number of n-poles) 
of standard
algebraical problems and represent all dynamical variables as expansion in the
bases of maximally localized in phase space functions (wavelet bases).   
Wavelet analysis is a relatively novel set of mathematical
methods, which gives us a possibility to work with well-localized bases in
functional spaces and gives for the general type of operators (differential,
integral, pseudodifferential) in such bases the maximum sparse forms. 
Our approach in this paper is based on the generalization 
of variational-wavelet 
approach from [1]-[8],
which allows us to consider not only polynomial but rational type of 
nonlinearities [9]. 
The solution has the following form
\begin{equation}\label{eq:z}
z(t)=z_N^{slow}(t)+\sum_{j\geq N}z_j(\omega_jt), \quad \omega_j\sim 2^j
\end{equation}
which corresponds to the full multiresolution expansion in all time 
scales.
Formula (\ref{eq:z}) gives us expansion into a slow part $z_N^{slow}$
and fast oscillating parts for arbitrary N. So, we may move
from coarse scales of resolution to the 
finest one for obtaining more detailed information about our dynamical process.
The first term in the RHS of equation (1) corresponds on the global level
of function space decomposition to  resolution space and the second one
to detail space. In this way we give contribution to our full solution
from each scale of resolution or each time scale.
The same is correct for the contribution to power spectral density
(energy spectrum): we can take into account contributions from each
level/scale of resolution.
Starting in part 2 from Hamiltonian of orbital motion in magnetic lattice
with additional kicks terms, we consider
in part 3 variational formulation for dynamical system with 
rational nonlinearities
and 
construct via multiresolution analysis 
explicit representation for all dynamical variables in the base of
compactly supported wavelets.

\section{PARTICLE IN THE MULTIPOLAR FIELD}

The magnetic vector potential of a magnet with $2n$ poles in Cartesian
coordinates is
\begin{equation}
A=\sum_n K_nf_n(x,y),
\end{equation}
where $ f_n$ is a homogeneous function of  $x$ and $y$ of order $n$.
The real and imaginary parts of binomial expansion of
\begin{equation}
f_n(x,y)=(x+iy)^n
\end{equation}
correspond to regular and skew multipoles. The cases $n=2$ to $n=5$
correspond to low-order multipoles: quadrupole, sextupole, octupole, decapole.
The corresponding Hamiltonian ([10] for designation):
\begin{eqnarray}\label{eq:ham}
&&H(x,p_x,y,p_y,s)=\frac{p_x^2+p_y^2}{2}+\nonumber\\
&&\left(\frac{1}{\rho(s)^2}-k_1(s)\right)
\cdot\frac{x^2}{2}+k_1(s)\frac{y^2}{2}\\
&&-{\cal R}e\left[\sum_{n\geq 2}
\frac{k_n(s)+ij_n(s)}{(n+1)!}\cdot(x+iy)^{(n+1)}\right]\nonumber
\end{eqnarray}
Then we may take into account arbitrary but finite number of terms in expansion
of RHS of Hamiltonian (\ref{eq:ham}) and
from our point of view the corresponding Hamiltonian equations of motions are
not more than nonlinear ordinary differential equations with polynomial
nonlinearities and variable coefficients.
Also we may add the terms corresponding to kick type contributions 
of rf-cavity:
\begin{eqnarray}
A_\tau=-\frac{L}{2\pi k}\cdot V_0\cdot \cos\big(k\frac{2\pi}
 {L}\tau\big)\cdot\delta(s-s_0)
\end{eqnarray}
or localized cavity $V(s)=V_0\cdot \delta_p(s-s_0)$ with $\delta_p(s-s_0)=
\sum^{n=+\infty}_{n=-\infty}\delta(s-(s_0+n\cdot L))$
at position $s_0$.  
Fig.1 and Fig.2  present finite kick term model and the
corresponding multiresolution representation on each level of resolution.
\begin{figure} [htb] 
\centering
\includegraphics*[width=60mm]{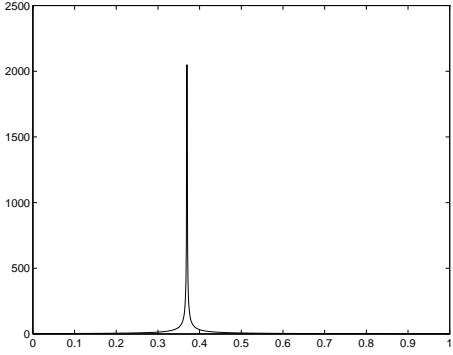}
\caption{Finite kick model.}
\end{figure}
\begin{figure} [htb] 
\centering
\includegraphics*[width=60mm]{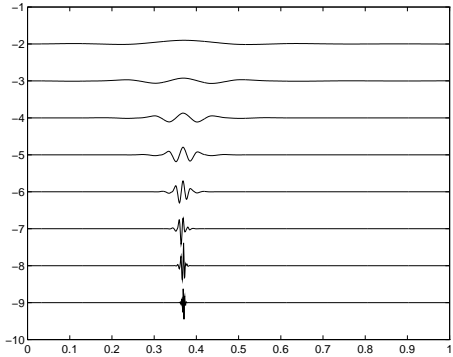}
\caption{Multiresolution representation of kick.}
\end{figure}

\section{RATIONAL DYNAMICS}

The first main part of our consideration is some variational approach
to this problem, which reduces initial problem to the problem of
solution of functional equations at the first stage and some
algebraical problems at the second stage.
We have the solution in a compactly
supported wavelet basis.
Multiresolution expansion is the second main part of our construction.
The solution is parameterized by solutions of two reduced algebraical
problems, one is nonlinear and the second are some linear
problems, which are obtained from one of the next wavelet
constructions:  the method of Connection
Coefficients (CC), Stationary Subdivision Schemes (SSS).

\subsection{Variational Method}
Our problems may be formulated as the systems of ordinary differential
equations
\begin{eqnarray}\label{eq:pol0}
& & Q_i(x)\frac{\ud x_i}{\ud t}=P_i(x,t),\quad x=(x_1,..., x_n),\\
& &i=1,...,n, \quad
 \max_i  deg \ P_i=p, \quad \max_i deg \  Q_i=q \nonumber
\end{eqnarray}
with fixed initial conditions $x_i(0)$, where $P_i, Q_i$ are not more
than polynomial functions of dynamical variables $x_j$
and  have arbitrary dependence of time. Because of time dilation
we can consider  only next time interval: $0\leq t\leq 1$.
 Let us consider a set of functions
\begin{eqnarray}
 \Phi_i(t)=x_i\frac{\ud}{\ud t}(Q_i y_i)+P_iy_i
\end{eqnarray}
and a set of functionals
\begin{eqnarray}
F_i(x)=\int_0^1\Phi_i (t)dt-Q_ix_iy_i\mid^1_0,
\end{eqnarray}
where $y_i(t) \ (y_i(0)=0)$ are dual (variational) variables.
It is obvious that the initial system  and the system
\begin{equation}\label{eq:veq}
F_i(x)=0
\end{equation}
are equivalent.
Of course, we consider such $Q_i(x)$ which do not lead to the singular
problem with $Q_i(x)$, when $t=0$ or $t=1$, i.e. $Q_i(x(0)), Q_i(x(1))\neq\infty$.

Now we consider formal expansions for $x_i, y_i$:
\begin{eqnarray}\label{eq:pol1}
x_i(t)=x_i(0)+\sum_k\lambda_i^k\varphi_k(t)\quad
y_j(t)=\sum_r \eta_j^r\varphi_r(t),
\end{eqnarray}
where $\varphi_k(t)$ are useful basis functions of  some functional 
space ($L^2, L^p$, Sobolev, etc) corresponding to concrete 
problem and
 because of initial conditions we need only $\varphi_k(0)=0$, $r=1,...,N, \quad i=1,...,n,$
\begin{equation}\label{eq:lambda}
\lambda=\{\lambda_i\}=\{\lambda^r_i\}=(\lambda_i^1, \lambda_i^2,...,\lambda_i^N), 
\end{equation}
 where the lower index i corresponds to 
expansion of dynamical variable with index i, i.e. $x_i$ and the upper index $r$
corresponds to the numbers of terms in the expansion of dynamical variables in the 
formal series.
Then we put (\ref{eq:pol1}) into the functional equations (\ref{eq:veq}) and as result
we have the following reduced algebraical system
of equations on the set of unknown coefficients $\lambda_i^k$ of
expansions (\ref{eq:pol1}):
\begin{eqnarray}\label{eq:pol2}
L(Q_{ij},\lambda,\alpha_I)=M(P_{ij},\lambda,\beta_J),
\end{eqnarray}
where operators L and M are algebraization of RHS and LHS of initial problem 
(\ref{eq:pol0}), where $\lambda$ (\ref{eq:lambda}) are unknowns of reduced system
of algebraical equations (RSAE)(\ref{eq:pol2}).

$Q_{ij}$ are coefficients (with possible time dependence) of LHS of initial 
system of differential equations (\ref{eq:pol0}) and as consequence are coefficients
of RSAE.

 $P_{ij}$ are coefficients (with possible time dependence) of RHS 
of initial system of differential equations (\ref{eq:pol0}) and as consequence 
are coefficients of RSAE.

$I=(i_1,...,i_{q+2}), \ J=(j_1,...,j_{p+1})$ are multiindexes, by which are 
labelled $\alpha_I$ and $\beta_I$ --- other coefficients of RSAE (\ref{eq:pol2}):
\begin{equation}\label{eq:beta}
\beta_J=\{\beta_{j_1...j_{p+1}}\}=\int\prod_{1\leq j_k\leq p+1}\varphi_{j_k},
\end{equation}
where p is the degree of polinomial operator P (\ref{eq:pol0})
\begin{equation}\label{eq:alpha}
\alpha_I=\{\alpha_{i_1}...\alpha_{i_{q+2}}\}=\sum_{i_1,...,i_{q+2}}\int
\varphi_{i_1}...\dot{\varphi_{i_s}}...\varphi_{i_{q+2}},
\end{equation}
where q is the degree of polynomial operator Q (\ref{eq:pol0}), 
$i_\ell=(1,...,q+2)$, $\dot{\varphi_{i_s}}=\ud\varphi_{i_s}/\ud t$.

Now, when we solve RSAE (\ref{eq:pol2}) and determine
 unknown coefficients from formal expansion (\ref{eq:pol1}) we therefore
obtain the solution of our initial problem.
It should be noted if we consider only truncated expansion (\ref{eq:pol1}) with N terms
then we have from (\ref{eq:pol2}) the system of $N\times n$ algebraical equations 
with degree $\ell=max\{p,q\}$
and the degree of this algebraical system coincides
 with degree of initial differential system.
So, we have the solution of the initial nonlinear
(rational) problem  in the form
\begin{eqnarray}\label{eq:pol3}
x_i(t)=x_i(0)+\sum_{k=1}^N\lambda_i^k X_k(t),
\end{eqnarray}
where coefficients $\lambda_i^k$ are roots of the corresponding
reduced algebraical (polynomial) problem RSAE (\ref{eq:pol2}).
Consequently, we have a parametrization of solution of initial problem
by solution of reduced algebraical problem (\ref{eq:pol2}).
The first main problem is a problem of
 computations of coefficients $\alpha_I$ (\ref{eq:alpha}), $\beta_J$ 
(\ref{eq:beta}) of reduced algebraical
system.
These problems may be explicitly solved in wavelet approach.

Next we consider the  construction  of explicit time
solution for our problem. The obtained solutions are given
in the form (\ref{eq:pol3}),
where
$X_k(t)$ are basis functions and
  $\lambda_k^i$ are roots of reduced
 system of equations.  In our case $X_k(t)$
are obtained via multiresolution expansions and represented by
 compactly supported wavelets and $\lambda_k^i$ are the roots of
corresponding general polynomial  system (\ref{eq:pol2})  with coefficients, which
are given by CC or SSS  constructions.  According to the
        variational method   to  give the reduction from
differential to algebraical system of equations we need compute
the objects $\alpha_I$ and $\beta_J$ [1],[9].

Our constructions are based on multiresolution app\-ro\-ach. Because affine
group of translation and dilations is inside the approach, this
method resembles the action of a microscope. We have contribution to
final result from each scale of resolution from the whole
infinite scale of spaces. More exactly, the closed subspace
$V_j (j\in {\bf Z})$ corresponds to  level j of resolution, or to scale j.
We consider  a multiresolution analysis of $L^2 ({\bf R}^n)$
(of course, we may consider any different functional space)
which is a sequence of increasing closed subspaces $V_j$:
\begin{equation}
...V_{-2}\subset V_{-1}\subset V_0\subset V_{1}\subset V_{2}\subset ...
\end{equation}
satisfying the following properties:
\begin{eqnarray}
&&\displaystyle\bigcap_{j\in{\bf Z}}V_j=0,\quad
\overline{\displaystyle\bigcup_{j\in{\bf Z}}}V_j=L^2({\bf R}^n),\nonumber
\end{eqnarray}
On Fig.3 we present contributions to solution 
of initial problem  
from first 5 scales or levels of resolution.
\begin{figure}
\centering
\includegraphics*[width=60mm]{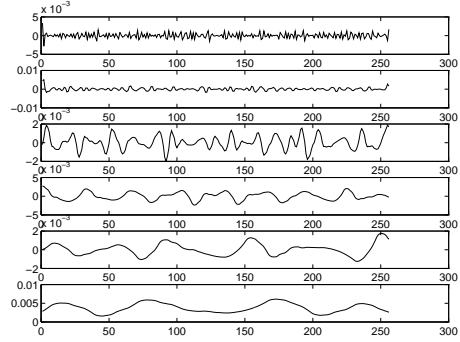}
\caption{Contributions to approximation: from scale $2^1$ to $2^5$.}
\end{figure}

We would like to thank Professor
James B. Rosenzweig and Mrs. Melinda Laraneta for
nice hospitality, help and support during UCLA ICFA Workshop.

\end{document}